\documentclass[aps,preprint,amsmath,amssymb,showpacs]{revtex4-1}
\usepackage{graphicx}

\begin{document}
\title
{\Large {\bf CP violation in neutrino--photon scattering}}

\author{ I. Alikhanov}
\email[]{E-mail: ialspbu@gmail.com}

\affiliation{Institute for Nuclear Research of the Russian Academy of Sciences,
60-th October Anniversary pr. 7a, Moscow 117312, Russia
}


\begin{abstract}
It is shown that effects of CP violation arise in neutrino--photon scattering. Several CP-violating $\nu\gamma$~reactions are considered within the Standard Model and predictions for some observables are made.  It is emphasized that neutrino--photon scattering may provide an experimental tool for testing the scale dependence of CP violation.

\end{abstract}
\maketitle %

\section{Introduction}
Even if almost 50 years have passed since the revolutionary discovery of the $K_L\rightarrow\pi\pi$ mode~\cite{cp1964}, CP-violating processes observed to date are still limited by meson weak decays. Moreover, CP-violating phenomena outside the kaon physics are experimentally  established only in the last decade at the $B$ factories~\cite{b_dec1,b_dec2,b_dec3,b_dec4,b_dec5,b_dec6,b_dec7}. Evidence for CP violation in neutral $D$ meson decays has also been achieved recently, but at a level lower than $5\sigma$~\cite{d_dec}. 
Other experimental efforts now being initiated to test CP symmetry like searching for the electric dipole moments of neutrons, electrons and nuclei as well as studying the decays of tau leptons reveal no CP violation.

Meanwhile, as noted by Sakharov in 1967~\cite{sakharov}, CP violation is a necessary condition for the appearance of the asymmetry between matter and antimatter in the universe. In spite of the impressive success demonstrated by the Standard Model in describing CP violation through the Cabibbo--Kobayashi--Maskawa (CKM) mechanism~\cite{ckm}, it fails by several orders of magnitude to account for the observed asymmetry. This is a challenging issue occupying a very unique  place in particle physics whose resolution  is contingent on  further endeavors in this field. 
There are several processes that are probably to play an important role in future CP studies as production and decay of the top quark and neutrino oscillations~\cite{cp_book2009}. 

Though neutrinos are generally considered to be weakly interacting particles, neutrino--photon
interactions should not be ignored or confined to discussions of loop effects in scattering, or generating
neutrino magnetic moments~\cite{seckel}. In some cases $\nu\gamma$ reactions already at tree level
are competitive with traditional neutrino reactions, and even may be dominant. An intuitively understandable view of how a neutrino interacts with the photon is provided by the parton model~\cite{mine1,mine2}. This paper demonstrates that within the Standard Model CP violation arises in $\nu\gamma$ reactions producing non-CP eigenstates (like $K_L$ and $K_S$). Several of such reactions are studied and predictions for some CP-violating observables are made.  

The purpose of this paper is to show that neutrino--photon scattering may also provide another tool for exploring CP violation in a wide energy range. 
The idea is based on the following. The CP transformation interchanges $\nu\gamma\leftrightarrow\bar\nu\gamma$. If CP is violated,  the $\nu\gamma$ scattering probability (or the total cross section) will differ from the  $\bar\nu\gamma$ one at the same kinematical conditions. In other words, $\nu$ and its antiparticle $\bar\nu$ will interact differently with the photon. Thus, by measuring the fluxes of neutrinos and antineutrinos sent through a field of photons, one can  determine the total CP-violating effect caused by all $\nu\gamma$ reactions allowed at the given energy, including the elastic channels.

\section{Exclusive production of $K_{S}$ and $K_L$ in neutrino--photon scattering}
As an example showing that CP can be violated in $\nu\gamma$ reactions, let us consider exclusive production of $K_{S}$ and $K_L$ in neutrino--photon scattering.

It has been experimentally established that the kaons $K_S$ and $K_L$ are mixtures of CP even and CP odd states~\cite{cp1964} and can be represented as a superposition of the $K^0$ and $\bar K^0$ mesons with definite strangeness~\cite{rev1996}:

\begin{equation}
|K_S\rangle=\frac{1}{\sqrt{2(1+|\tilde{\epsilon}|^2)}}\left((1+\tilde{\epsilon})|K^0\rangle+(1-\tilde{\epsilon})|\bar K^0\rangle\right),
\label{ks}
\end{equation} 

\begin{equation}
|K_L\rangle=\frac{1}{\sqrt{2(1+|\tilde{\epsilon}|^2)}}\left((1+\tilde{\epsilon})|K^0\rangle-(1-\tilde{\epsilon})|\bar K^0\rangle\right),
\label{kl}
\end{equation} 

where $\tilde\epsilon$ is a complex parameter.

Consider the following semileptonic decays of $K_{L,S}$:

\begin{equation}
K_{L,S}\put(1,5){\vector(4,1){25}}\put(1,2){\vector(4,-1){25}}\put(30,10){$\pi^-l^+\nu_l$}\put(30,-10){$\pi^+l^-\bar\nu_l$.}\label{decay1}
\end{equation} 

Here $l=e, \mu$. 

The corresponding decay amplitudes can be expressed in terms of the amplitude of the process $K^0 \rightarrow \pi^-l^+\nu_l$ by using~(\ref{ks})--(\ref{kl}) and taking into account the selection rule $\Delta S=\Delta Q$~\cite{rev1996}:

\begin{equation}
A(K_{L} \rightarrow \pi^-l^+\nu_l)=A(K_{S} \rightarrow \pi^-l^+\nu_l)\\=\frac{1+\tilde\epsilon}{\sqrt{2(1+|\tilde{\epsilon}|^2)}}A(K^0 \rightarrow \pi^-l^+\nu_l),\label{ampl1}
\end{equation} 

\begin{equation}
-A(K_{L} \rightarrow \pi^+l^-\bar\nu_l)=A(K_{S} \rightarrow \pi^+l^-\bar\nu_l)\\=\frac{1-\tilde\epsilon}{\sqrt{2(1+|\tilde{\epsilon}|^2)}}A^*(K^0 \rightarrow \pi^-l^+\nu_l).\label{ampl2}
\end{equation} 

Note that $A (\bar K^0 \rightarrow \pi^+l^-\bar\nu_l)=A^*(K^0 \rightarrow \pi^-l^+\nu_l)$. In addition, the selection rule $\Delta S=\Delta Q$ implies that 
$A(K^0 \rightarrow \pi^+l^-\bar\nu_l)=A(\bar K^0 \rightarrow \pi^-l^+\nu_l)=0$.

Let us turn now to neutrino--photon scattering. The CP transformation yields

\begin{equation}
\nu_l\gamma\overset{\text{CP}}{\Longleftrightarrow}\bar\nu_l\gamma.
\end{equation} 

If CP symmetry is conserved, then the cross sections of $\nu_l\gamma$ and $\bar\nu_l\gamma$ reactions at the same kinematical conditions must be equal, i.e.

\begin{equation}
\sigma(\nu_l\gamma\rightarrow X)=\sigma(\bar\nu_l\gamma\rightarrow X^{\prime}).\label{sigma_x}
\end{equation} 

It will be shown below that the Standard Model predicts violation of~(\ref{sigma_x}).

Let us analyze the following reactions:

\begin{equation}
\nu_l\gamma\rightarrow l^-\pi^+K_{L,S}, \label{nu1}
\end{equation} 

\begin{equation}
\bar\nu_l\gamma\rightarrow l^+\pi^-K_{L,S}.\label{nu2}
\end{equation}

\begin{figure}
\centering
\resizebox{0.3\textwidth}{!}{%
\includegraphics{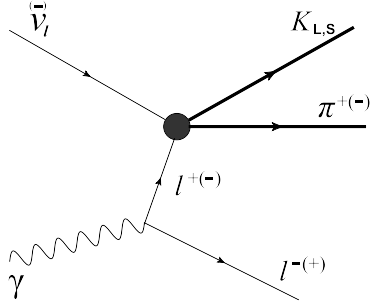}
}\caption{Production of the $l^-\pi^+K_{L,S}$ ($l^+\pi^-K_{L,S}$) final state in $\nu_l\gamma$ ($\bar\nu_l\gamma$) scattering.}
\label{fig1}
\end{figure}

For this purpose, it is very convenient to use the \newline equivalent-lepton approximation \cite{zerwas} within which the incident neutrino (antineutrino) interacts with the corresponding charged lepton resulting from photon splitting into $l^+l^-$ pairs. This mechanism is illustrated by the diagram shown in Fig.~\ref{fig1}.
It should be emphasized that now the involved neutrino flavors are not limited only by $\nu_e$ and $\nu_{\mu}$ as in~(\ref{decay1}) but may also be $\nu_{\tau}$.

Thus, to find the cross sections of (\ref{nu1}) and (\ref{nu2}) one needs to know the cross sections of the following subprocesses:

\begin{equation}
\nu_ll^+\rightarrow \pi^+K_{L,S}\label{nu1_sub},
\end{equation} 

\begin{equation}
\bar\nu_ll^-\rightarrow \pi^-K_{L,S}.\label{nu2_sub}
\end{equation} 

and convolute them with the equivalent-lepton spectrum of the photon.

Noting that the reactions~(\ref{nu1_sub}) and (\ref{nu2_sub}) are related by crossing symmetry to the decays~(\ref{decay1}) one can derive the corresponding amplitudes from~(\ref{ampl1}) and (\ref{ampl2}), so that:  

\begin{equation}
A(\nu_ll^+\rightarrow \pi^+K_L)=A(\nu_ll^+\rightarrow \pi^+K_S)\\=\frac{1+\tilde\epsilon}{\sqrt{2(1+|\tilde{\epsilon}|^2)}}A(\nu_ll^+\rightarrow \pi^+K^0),\label{ampl3}
\end{equation} 

\begin{equation}
-A(\bar\nu_ll^-\rightarrow \pi^-K_L)=A(\bar\nu_ll^-\rightarrow \pi^-K_S)\\=\frac{1-\tilde\epsilon}{\sqrt{2(1+|\tilde{\epsilon}|^2)}}A^*(\nu_ll^+\rightarrow \pi^+K^0).\label{ampl4}
\end{equation}

From~(\ref{ampl3}) and (\ref{ampl4}) it is obvious that the cross sections of~(\ref{nu1_sub}) and (\ref{nu2_sub}) can be written as

\begin{equation}
\sigma(\nu_ll^+\rightarrow \pi^+K_{L,S})=\frac{|1+\tilde\epsilon|^2}{2(1+|\tilde{\epsilon}|^2)}\hat\sigma(s),\label{cross_sub1}
\end{equation} 

\begin{equation}
\sigma(\bar\nu_ll^-\rightarrow \pi^-K_{L,S})=\frac{|1-\tilde\epsilon|^2}{2(1+|\tilde{\epsilon}|^2)}\hat\sigma(s), \label{cross_sub2}
\end{equation} 

where $\hat\sigma(s)$ represents the dependence of the cross sections on the total center-of-mass (cms) energy $s$.

The standard convolution of~(\ref{cross_sub1}) and (\ref{cross_sub2}) with the charged lepton distributions in the photon gives the sought-for cross sections:

\begin{equation}
\sigma(\nu_l\gamma\rightarrow l^-\pi^+K_{L,S})=\frac{|1+\tilde\epsilon|^2}{2(1+|\tilde{\epsilon}|^2)}\int\hat\sigma(xs)f_{\gamma}^{l^+}(x,s)dx,\label{cross_11}
\end{equation}

\begin{equation}
\sigma(\bar\nu_l\gamma\rightarrow l^+\pi^-K_{L,S})=\frac{|1-\tilde\epsilon|^2}{2(1+|\tilde{\epsilon}|^2)}\int\hat\sigma(xs)f_{\gamma}^{l^-}(x,s)dx.\label{cross_22}
\end{equation}

Here $f_{\gamma}^{l^\pm}(x,s)$ is the probability density of finding a charged lepton $l^{\pm}$ in the photon with a fraction $x$ of the parent photon's momentum (in other words, the equivalent-lepton spectrum of the photon). 

It is crucial that the probability of finding $l^+$ in the photon is equal to that of finding $l^-$ for any fixed $x$ and $s$, i.e. 

\begin{equation}
f_{\gamma}^{l^+}(x,s)=f_{\gamma}^{l^-}(x,s),\label{equal} 
\end{equation}

otherwise CP symmetry would be violated even in electromagnetic interactions.

Taking into account~(\ref{equal}) one can see that~(\ref{cross_11}) and (\ref{cross_22}) do not satisfy the relation~(\ref{sigma_x}) and CP symmetry is violated.
One can introduce a measure of CP violation in these reactions as

\begin{equation}
\delta_{L,S}=\frac{\sigma(\nu_l\gamma\rightarrow l^-\pi^+K_{L,S})-\sigma(\bar\nu_l\gamma\rightarrow l^+\pi^-K_{L,S})}{\sigma(\nu_l\gamma\rightarrow l^-\pi^+K_{L,S})+\sigma(\bar\nu_l\gamma\rightarrow l^+\pi^-K_{L,S})}\\=\frac{2{\cal R}e(\bar\epsilon)}{1+|\bar\epsilon|^2}.\label{assym}
\end{equation}

The quantity $\delta_{L,S}$ exactly coincides with the parameter $\delta_L$  used in  studies of  the CP-violating semileptonic $K_L$ decays~\cite{rev1996} and whose value is known from the experiment~\cite{pdg2012}. So, one makes the following prediction: 

\begin{equation}
\delta_{L,S}=\delta_L=(0.332\pm0.006)\%\label{predict}.
\end{equation}

The production of $l^-$ (or $\pi^+$) in the reaction~(\ref{nu1}) turns out to be more probable than that of $l^+$ ($\pi^-$) in~(\ref{nu2}). In other words, CP violation leads to an asymmetry between  $l^-$ and $l^+$ ($\pi^+$ and $\pi^-$).

\begin{figure}
\centering
\resizebox{0.3\textwidth}{!}{%
\includegraphics{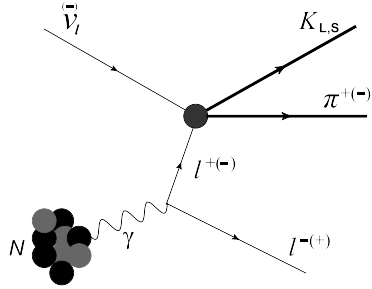}
}\caption{Production of the $l^-\pi^+K_{L,S}$ ($l^+\pi^-K_{L,S}$) final state in neutrino--nucleus (antineutrino--nucleus) scattering.}
\label{fig2}
\end{figure}

In addition to reactions with real photons, the asymmetry $\delta_{L,S}$ can be measured in interaction of neutrinos (antineutrinos) with an external electromagnetic field. For example, this can be done in experiments on neutrino--nucleus scattering, where the neutrinos interact with the virtual photons of the Coulomb field of the nucleus as displayed schematically in Fig.~\ref{fig2}. Again, the corresponding cross sections can be found by the  convolution of~(\ref{cross_11}) and (\ref{cross_22}) with the equivalent-photon spectrum of the nucleus. One can easily show that $\delta_{L,S}$ for this case is also equal to~(\ref{assym}), i.e.

\begin{equation}
\frac{\sigma(\nu_lN\rightarrow l^-\pi^+K_{L,S}N)-\sigma(\bar\nu_lN\rightarrow l^+\pi^-K_{L,S}N)}{\sigma(\nu_lN\rightarrow l^-\pi^+K_{L,S}N)+\sigma(\bar\nu_lN\rightarrow l^+\pi^-K_{L,S}N)}\\=\frac{2{\cal R}e(\bar\epsilon)}{1+|\bar\epsilon|^2}.\label{assym2}
\end{equation}


Though the quantity~(\ref{assym}) is found within the Standard Model, there are however at least two reasons that enable one to expect deviations from this prediction and thus make the considered reactions nontrivial for testing CP violation in the Standard Model: 

1) The parameter  $\tilde\epsilon$ being actually a part of the reaction amplitudes may depend on the cms energy $s$ and the momentum transfer squared $t$, i.e.

\begin{equation}
\tilde\epsilon=\tilde\epsilon(s,t).
\end{equation}

For example, within the Standard Model and the CKM formalism one has

\begin{equation}
|\tilde\epsilon|=\frac{G_F^2f^2_Km^2_cm_K}{6\sqrt{2}\pi^2\Delta m}A^2\lambda^6\eta B_K\left(-g(x_c)+A^2\lambda^4(1-\rho)\frac{x_t^2}{x^2_c}g(x_t)+h(x_c,x_t)\right),\label{sm_epsilon}
\end{equation}

where $G_F$ is the weak coupling constant, $m_K$, $f_K$, and $B_K$ are the mass, decay constant and bag factor of the kaon, $m_c$ and $m_t$ are the charm and top quark masses, $A$, $\lambda$, $\eta$ and $\rho$ are the CKM parameters in the Wolfenstein representation, $g(x_c)$ and $h(x_c, x_t)$ are kinematic functions of $x_{c,t}=m^2_{c,t}/m^2_W$ with $m_W$ being the mass of $W$ boson, $\Delta m$ is the real part of the $\bar K^0$--$K^0$ transition matrix element. For details of deriving~(\ref{sm_epsilon}) see, e.g.,~\cite{ho-kim,cp_book1999}.

The functions $g(x_c)$ and $h(x_c, x_t)$ describe the loop contributions from the $c$ and $t$ quarks including QCD corrections, the factors $f_K$ and $B_K$ are also parametrize the non-perturbative strong corrections, so that these quantities may be obviously scale dependent. 

Meanwhile, the experimental value $\delta_L=(0.332\pm0.006)\%$ measured in the semileptonic $K_L$ decays alwyas corresponds to some fixed $s$ and $t$ whose domain of variation is restricted by the mass of $K_L$. Unlike these decays, the reactions~(\ref{nu1})--(\ref{nu2}) will probe $\tilde\epsilon$ at higher values of $s$ and $t$ and in this case $\delta_{L,S}$ may differ from~$\delta_L$. It should be also emphasized that the $K_L$ mesons used to measure $\delta_L$ are initially produced in the $K^0$ or $\bar K^0$ states with defenite strangeness in strong reactions  such as $p\bar p\rightarrow K^-\pi^+K^0$ or $p\bar p\rightarrow K^+\pi^-\bar K^0$. In contrast, the weak interaction responsible for the reactions~(\ref{nu1}) and (\ref{nu2}) does not conserve the strangeness and accommodates the possibility of direct production of $K_L$ and $K_S$ mesons which have no definite strangeness as implied by the parametrization of the interaction given by~(\ref{ampl3})--(\ref{ampl4}). 

2) In obtaining the amplitudes~(\ref{ampl3})--(\ref{ampl4}), the principle of detailed balance is implicitly used. The latter is related to time-reversal invariance. Since we deal with CP-nonconserving processes, the CPT theorem requires T violation as well and therefore the true amplitudes may deviate from those given by~(\ref{ampl3})--(\ref{ampl4}).

\section{Discussion and conclusions}

The CP transformation interchanges $\nu\gamma\leftrightarrow\bar\nu\gamma$. Therefore, the cross sections of $\nu\gamma$ and $\bar\nu\gamma$ reactions at the same kinematical conditions must be equal if CP is conserved.

It is shown that the Standard Model predicts CP violation in neutrino--photon scattering. In particular, CP-violating effects can manifest themselves in the reactions $\nu_l\gamma\rightarrow l^-\pi^+K_{L,S}$ and $\bar\nu_l\gamma\rightarrow l^+\pi^-K_{L,S}$ in the form of an asymmetry between the final leptons $l^-$ and $l^+$ (or pions $\pi^+$ and $\pi^-$). This asymmetry is found to exactly coincide with the parameter $\delta_L$ measured in the semileptonic $K_L\rightarrow l^+\nu_l\pi^-$  and $K_L\rightarrow l^-\bar\nu_l\pi^+$ decays.

There may be more complicated CP-violating processes in which neutrino--photon scattering proceeds in deep inelastic regime and the neutral kaons are produced along with a multiparticle final state $X$: $\nu_l\gamma\rightarrow l^-K_{L,S}X$, $\bar\nu_l\gamma\rightarrow l^+K_{L,S}X^{\prime}\label{nu22}$.
Apart from the reactions with the neutral kaons, the $l^+l^-$-asymmetry can also be searched for in production of other mesons in decays of which CP violation has been either observed or expected. For example, one can investigate  production of $D$ and $B$ mesons.
The CP-violating effects observed in $B$ decays are larger than in the kaon decays~\cite{pdg2012} and therefore one could expect that they are significant in reactions like $\nu_l\gamma\rightarrow l^-BX$ and $\bar\nu_l\gamma\rightarrow l^+\bar BX^{\prime}$ as well. 

Another possibility of observation of CP violation can be provided by experiments on neutrino--nucleus scattering, where the neutrino interacts with the virtual photons of the Coulomb field of a target nucleus. The charge asymmetry $\delta_{L,S}$ predicted for this case also coincides with the one measured in the semileptonic $K_L$ decays.

\begin{figure}
\centering
\resizebox{0.7\textwidth}{!}{%
\includegraphics{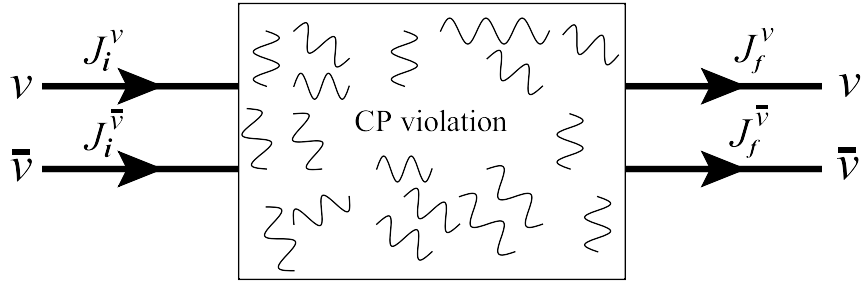}
}\caption{Experiment for testing CP violation. It measures the fluxes of neutrinos and antineutrinos sent through a field of photons after passing this field. The initial $\nu$ and $\bar\nu$ energy distributions ($J^{\nu}_i$, $J^{\bar\nu}_i$)  are identical up to a normalization factor.  CP violation will lead to the appearance of the difference between these distributions in the final state ($J^{\nu}_f$, $J^{\bar\nu}_f$) and the original identity will be lost. External electromagnetic fields, the Coulomb field of a nucleus, lasers can serve as the sources of the field of photons.}
\label{fig3}
\end{figure}

It should be emphasized that $\nu\gamma$ scattering may provide a tool for testing  CP violation without reference to any produced final state. Consider a hypothetical experiment in which $\nu$ and $\bar\nu$ beams are sent through an external electromagnetic field (the latter is nothing but a field of photons). This is illustrated in Fig.~\ref{fig3}. Assume that the initial energy distributions of the neutrinos $J^{\nu}_i$ and antineutrinos $J^{\bar\nu}_i$ are identical  up to a normalization factor. CP~violation implies that $\nu$ and $\bar\nu$ interact differently with photons. If CP violation occurs, these beams will behave differently while passing the field and the original identity between the energy distributions will disappear. By measuring the final neutrino fluxes one can judge whether CP is violated or not. 
Let $J^{\nu}_f$ and $J^{\bar\nu}_f$ be the energy distributions of the neutrinos and antineutrinos after passing the field of photons. Then the manifestation of CP violation can be formally expressed as 

\begin{equation}
\frac{J^{\nu}_i-J^{\nu}_f}{J^{\nu}_i}\neq\frac{J^{\bar\nu}_i-J^{\bar\nu}_f}{J^{\bar\nu}_i}. \label{nu111}
\end{equation} 

Such an experiment has a particular advantage that it determines the total effect caused by all CP-violating $\nu\gamma$-channels open at the given energy, including the elastic ones. 

It is interesting to speculate on the consequences of this effect for cosmology. For example, CP~violation might cause a $\nu_l\bar\nu_l$-asymmetry in cosmic rays with a  $\nu_l\bar\nu_l$-component  due to neutrinos (antineutrinos) can propagate cosmological distances while being bathed in the cosmic microwave background radiation and/or strong magnetic fields.

It is notable that neutrino--photon scattering allows to probe CP violation in a wide kinematical range, while the energy region covered by the experiments studying CP-nonconserving meson decays is restricted by the masses of the mesons.
\acknowledgements
This work was supported in part by the Russian Foundation for Basic Research (grant 11-02-12043), by the Program for Basic Research of the Presidium of the Russian Academy of Sciences "Fundamental Properties of Matter and Astrophysics" and by the Federal Target Program  of the Ministry of Education and Science of Russian Federation "Research and Development in Top Priority Spheres of Russian Scientific and Technological Complex for 2007-2013" (contract No. 16.518.11.7072).

\end{document}